\documentclass[]{raa}            

\usepackage{graphicx,times}             
\usepackage{natbib}
\usepackage{amssymb,amsmath}
\usepackage{graphicx}	
\usepackage{amsmath}	
\usepackage{graphicx,colortbl}
\usepackage{txfonts}
\usepackage{pdflscape,lscape}
\usepackage{graphicx}
\usepackage{rotating}
\usepackage{newtxtext,newtxmath}
\usepackage{tablefootnote}
\usepackage{soul}

\bibpunct{(}{)}{;}{a}{}{,}

\newcommand{\km}{${\rm km}\,{\rm s}^{-1}$}

\newcommand{\hi}{H\,{\sc i}}

\newcommand{\msolar}{${\rm M}_{\odot}$}

\newcolumntype{L}[1]{>{\raggedright\arraybackslash}p{#1}}
\newcolumntype{R}[1]{>{\raggedleft\arraybackslash}p{#1}}

\def\apjl{Astrophysical Journal Letters}
\def\aap{Astronomy and Astrophysics}

\def\apjs{The Astrophysical Journal Supplement Series}

\begin{document}

   \title{\textbf HI in high gas-phase metallicity dwarf galaxy WISEA J230615.06+143927.9
}
  \volnopage{Vol.0 (20xx) No.0, 000--000}      
   \setcounter{page}{1}          
 
   \author{Yan Guo
      \inst{1,2}
   \and C. Sengupta
      \inst{1}
   \and T. C. Scott
      \inst{3}
    \and P. Lagos
      \inst{3}
  \and Y. Luo
       \inst{1} 
 }

   \institute{Purple Mountain Observatory, Chinese Academy of Sciences, 10 Yuanhua Road, Nanjing,  Jiangsu 210023, China; {\it sengupta.chandreyee@gmail.com}\\
        \and
            School of Astronomy and Space Sciences, University of Science and Technology of China, 96 Jinzhai Road, Hefei, Anhui 230026, China\\
      \and
          Institute of Astrophysics and Space Sciences (IA Porto), Rua das Estrelas, 4150-762, Porto, Portugal\\  
             \vs\no
   {\small Received~~20xx month day; accepted~~20xx~~month day}}
   
\abstract{We present resolved GMRT \hi\ observations of the high gas-phase metallicity dwarf galaxy WISEA J230615.06+143927.9 (z = 0.005) (hereafter J2306) and investigate whether it could be a Tidal Dwarf Galaxy (TDG) candidate. TDGs are observed to have higher metallicities than normal dwarfs. J2306 has an unusual combination of  a blue g -- r colour of 0.23 mag, irregular optical morphology and high--metallicity  (12 + log(O/H) = 8.68$\pm$0.14), making it an interesting galaxy to study in more detail. We find J2306 to be an \hi\ rich galaxy with a large extended, unperturbed rotating \hi\ disk. Using our \hi\ data we estimated its dynamical mass and found the galaxy to be dark matter (DM) dominated within its \hi\ radius. The quantity of DM, inferred from its dynamical mass, appears to rule out J2306  as an evolved TDG. A wide area environment search reveals  J2306 to be isolated from any larger galaxies which could have been the source of its high gas metallicity. Additionally, the \hi\ morphology and kinematics of the galaxy show no indication of a recent merger to explain the high--metallicity. Further detailed optical spectroscopic observations of J2306 might provide an answer to how a seemingly ordinary irregular dwarf  galaxy achieved such a high level of metal enrichment.
}

   \authorrunning{Guo et al.}            
   \titlerunning{HI in metal rich dwarf galaxy}  

   \maketitle
%
%

\section{Introduction}           
\label{section_6dfgs_intro}

\begin{figure}

\begin{center}
\includegraphics[width=8.8cm]{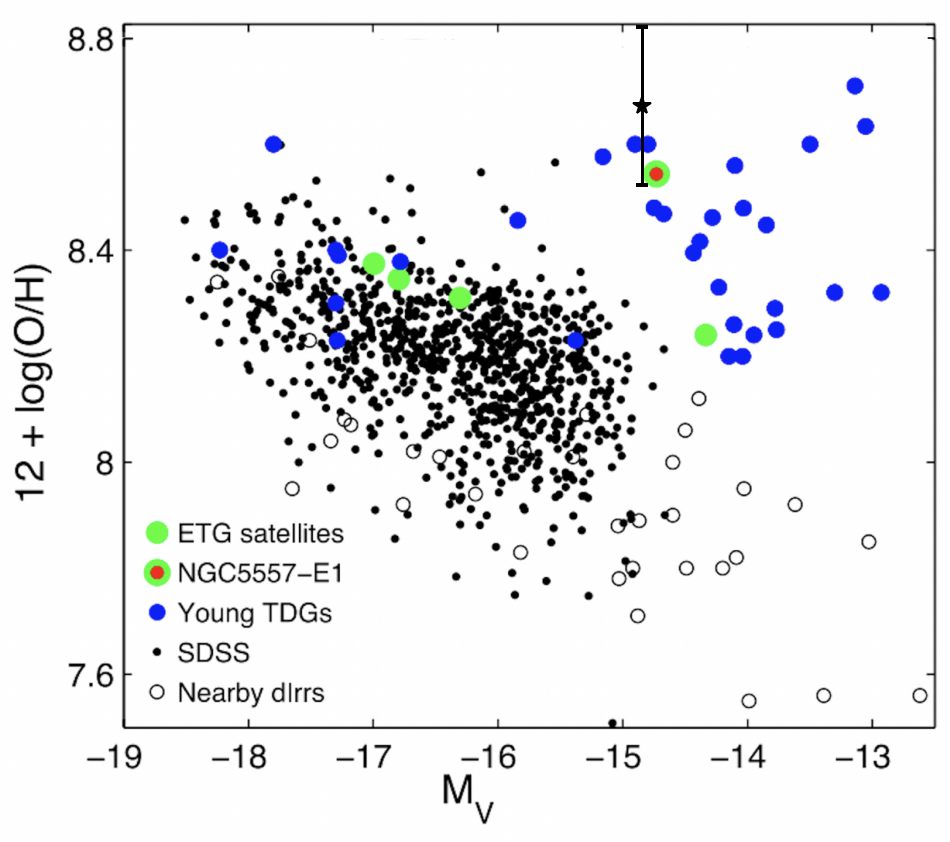} 
\caption{12 + log(O/H) abundances from selected samples of dwarf galaxies plotted against their V--band magnitude (M$_{V}$) \citep{duc2014}. The open circles are nearby dwarf irregulars from \citep{1995ApJ...445..642R, 2006ApJ...636..214V}, the filled black dots  represent normal SDSS dwarfs and the green filled circles and the NGC 5557-E1 are early-type galaxy satellites from \cite{duc2014}, the filled blue circles are  TDGs from \cite{2003A&A...397..545W, 2010AJ....140.2124B}. See figure 4 of \cite{duc2014} for more detail. J2306 is plotted with a black star to show its properties relative to other dwarf galaxies.}
\label{fig1}
 \end{center}
\end{figure}

\noindent  Extrapolating from the local universe,  low--mass dwarf galaxies are  understood to be the most ubiquitous galaxies in the universe \citep{dale09,mccon2012}, however, the local dwarfs remain relatively unexplored, because of the difficulty of observing such optically faint objects.  It is in this regime where cosmological predictions differ most from observations and the large scatter in the observed values of dwarf properties makes it difficult to discern the underlying trends in their properties.  Typically, dwarfs are dark matter (DM) dominated, faint irregular optical objects \citep[e.g.,][]{lagos2007}, with their observed DM potentials  being shallower than predicted by cosmological models \citep[e.g., ][]{Oh15}. Dwarf galaxies are also observed to have  signiﬁcantly sub--solar metallicities, 12 + log(O/H) = 7.4 to 7.9, occasionally reaching extremely low values \citep[][and references therein]{lee03,lagos2018}. \textcolor{black}{ \cite{tremonti} using Sloan Digital Sky Survey (SDSS) DR4 data, demonstrated the strong positive correlation between galaxy stellar mass and metallicty (12 + log(O/H)). The scatter in the relation is much larger in the dwarf mass range and  \cite{peeples08}, applying restrictive selection criteria to  the \cite{tremonti} sample, identified 41} dwarf galaxies \textcolor{black}{having}  metallicities between 8.6 $\le$ 12 + log(O/H) $\le$ 9.3, challenging the idea that dwarf galaxies are necessarily low--metallicity galaxies. Since the dwarf galaxy population is dominated by low--metallicity objects, the rarely found high--metallicity dwarfs are particularly interesting. While a handful of such high--metallicity dwarfs have been detected, their formation scenarios or physical properties are yet to be explored in detail. One possible formation scenario for such  high--metallicity dwarfs is  formation out of metal-- rich gas debris ejected during a tidal interaction between larger galaxies, at least one of which is gas rich \citep{duc99}. These tidal dwarf galaxies (TDGs), are usually observed to have higher metallicities than normal dwarfs of the same stellar mass, see blue points in Figure \ref{fig1} and \citep[e.g.][]{duc99,duc12,smith10,scott18}. Other possible formation scenarios may include accretion of tiny evolved early--type dwarf galaxies. In this paper, we present Giant Metrewave Radio Telescope (GMRT) \hi\ 21 cm imaging of the high--metallicity dwarf galaxy,  SDSS J230614.96+143926.7 also known as WISEA J230615.06+143927.9 (23h 06m 15.050s +14d39m27.50s), hereafter referred to as J2306, selected from the SDSS. The previous single dish \hi\ detection of the galaxy was with the Arecibo  telescope \citep[][AGC 332413]{haynes11}.  Being nearby and having a prior single dish \hi\ detection made J2306 a suitable candidate for a high--resolution \hi\ study.  More information on  the galaxy's properties  is provided in Section 2.1.  Our aim is to increase the understanding of  J2306's physical properties and investigate  possible formation scenarios. 

Using the  heliocentric optical velocity (1542 \km) of J2306  from NASA Extragalactic Database
(NED) and assuming H$_0$ = 68 \km/Mpc \citep{Planck-18}, we adopt a distance of 22.7 Mpc {\bf to} the galaxy. At this distance, the spatial scale is $\sim$ 6.4 kpc/arcmin. All $\alpha$ and $\delta$ positions referred
to throughout this paper are J2000.


\section{Data and Observations}
\subsection{The source properties and metallicity estimates}
\label{source_props}
\begin{figure}
\begin{center}
\includegraphics[width=8.8cm]{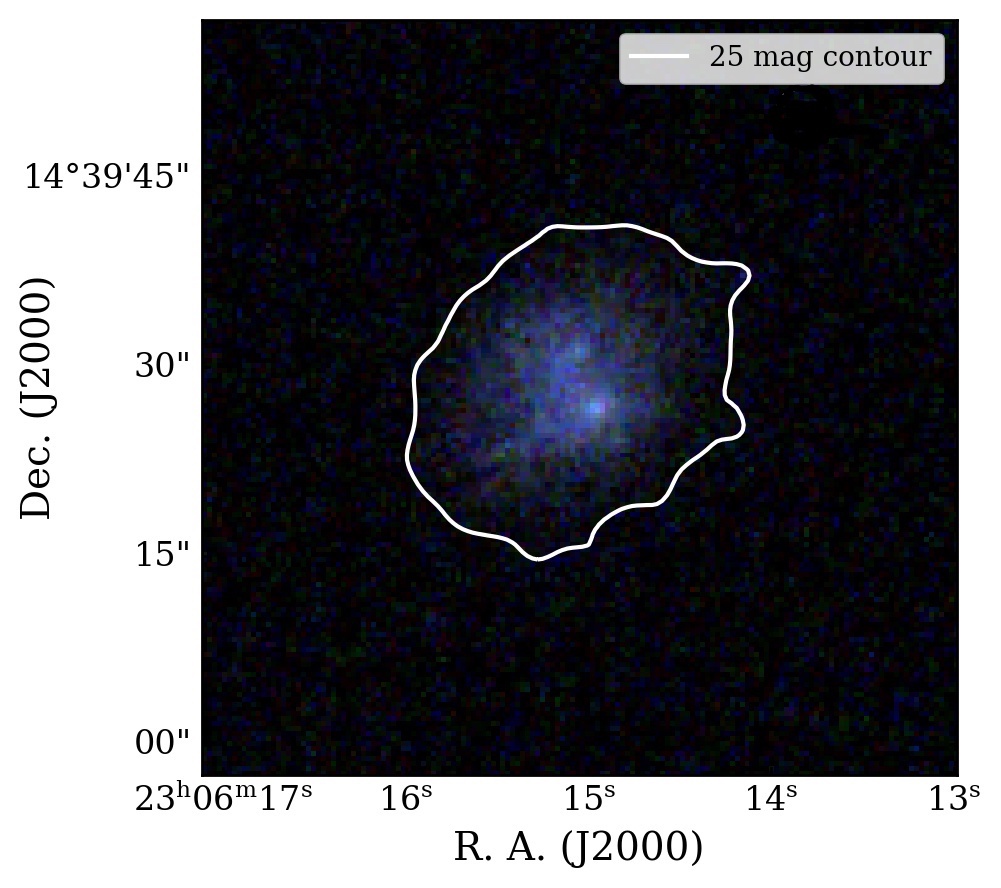} 
\caption{ SDSS g, r, i band composite image of J2306, with the 25th magnitude shown with a white contour} 
\label{fig-sdss}
 \end{center}
\end{figure}

J2306 is a small nearby dwarf galaxy with an SDSS g--band absolute magnitude = --14.83 mag, g -- r colour $\sim$ 0.23 mag and   its r--band D$_{25}$ diameter estimated from NED is $\sim$ 2.9 kpc.  The galaxy is not an early--type dwarf, as its optical morphology is non-compact and irregular.  Figure \ref{fig-sdss} shows the SDSS optical image of J2306. We used the relationship between photometric SDSS model magnitudes for selected colour band filters and stellar M/L ratio from \cite{bell03}, to estimate the galaxy's stellar mass. The SDSS g--band luminosity ($L_g$) and $g-r$ colour of the source were used to calculate  stellar mass using Equation \ref{eq:1}:  

\begin{equation}\label{eq:1}
    log(M/L_g) = -0.499+1.519\left(g-r\right)
\end{equation}

The stellar mass  (M$_*$) of J2306, calculated using equation \ref{eq:1} is 2.4 $\times$ 10$^7$ \msolar, confirming it as a low stellar mass dwarf. Typically, the distance to a galaxy can estimated from its recessional velocity which is related to the expansion rate of universe via the Hubble constant. However, for galaxies with recessional velocities $<$ 1500 \km\ peculiar motions along the line of sight due to local gravitational forces can add significant additional uncertainty  to the distance estimates. For J2306 adding or subtracting a typical group velocity dispersion (250 \km) to the measured recessional heliocentric velocity (1542 \km) changes the distance to the galaxy between 19.10 Mpc and 26.45 Mpc, which in turn implies M$_*$ = 1.7 $\times$ 10$^7$ \msolar\ and 3.2 $\times$ 10$^7$ \msolar, respectively. This shows that, unless J2306 has an extraordinarily high peculiar velocity, it is a low stellar mass dwarf with M$_*$ of the order 10$^7$ \msolar. J2306 is a faint galaxy that has an `unreliable photometry' flag in the SDSS database.  As a result of the uncertainties in both distance and photometry for J2306, we  treat  our stellar mass  calculation as only an order of magnitude estimate. 


\begin{figure}

\begin{center}
\includegraphics[width=15.0cm]{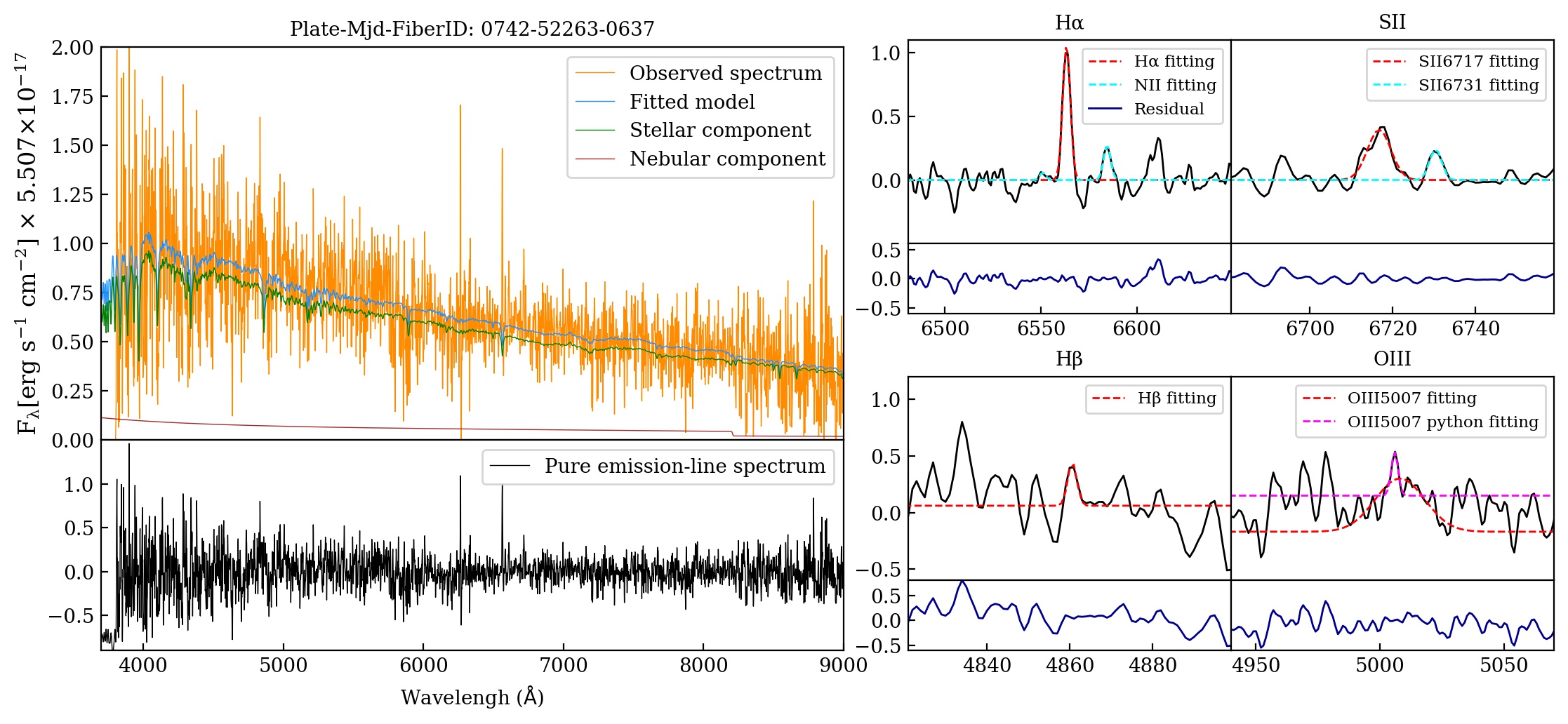} 
\caption{ The top left panel shows the SDSS spectrum of J2306 along with FADO fits. The bottom left panel shows the pure emission--line spectrum. The four panels to the right shows the FADO fits to individual lines and their residuals. Line names are shown on top of each panel. The [OIII] box has two different fits to the [OIII]$\lambda$5007 line. Due to poor SNR, the FADO fit (in red broken line) had failed. The line was then fitted using our own python code (shown in pink broken line).  The [OIII]$\lambda$4959 line is unresolved.} 
\label{fig-spec}
 \end{center}
\end{figure}

From J2306's SDSS spectrum (Figure \ref{fig-spec}), we estimated its gas phase metallicity. The spectrum's [OIII] $\lambda$4363 auroral emission line is weak and [O II] $\lambda$3727 is beyond the SDSS spectrometer’s wavelength range, so the T$_e$ method could not be used to calculate gas metallicity. Instead, we used the relation between metallicity and two strong line calibrators (N2 and O3N2) as prescribed in  \cite{denicolo02}, \cite{perez09}, \cite{PettiniPagel2004}, \cite{marino13}, \cite{bian18}, \cite{Naga06} and \cite{curti20} to estimate the 12 + log(O/H). 
We took the average value of  the 12 + log(O/H) derived from the N2 and O3N2 calibrations by each of the above sets of authors to calculate the final gas metallicity value of J2306. The complete list of the 12 + log(O/H)  calibrators from  the selected authors used to calculate the galaxy's metallicity is given in Table \ref{tab:tab1}.

\begin{table}
\centering
\caption{Metallicity for J2306 \textcolor{black}{using N2 calibrators from selected authors}.}
\label{tab:tab1}
\begin{tabular}{llcccccc}
\hline
Calibrator & Reference & 12 + log(O/H) \\
\hline
N2   & Bian 2018           & 8.69$\pm$0.20\\
N2   & Curti 2020          & 8.72$\pm$0.28 \\
N2   & Denicoló 2002       & 8.79$\pm$0.23\\
N2   & Marino 2013         & 8.53$\pm$0.15\\
N2   & Nagao 2006          & 9.01$\pm$0.33\\
N2   & Pérez--Montero 2009 & 8.71$\pm$0.25\\
N2   & Pettini 2004        & 8.68$\pm$0.74\\
\hline
mean N2 & &  8.73\\
std  N2 & &  0.13\\
\hline
O3N2 &  Bian2018           & 8.80$\pm$0.45\\
O3N2 & Curti2020           & 8.64$\pm$0.26\\
O3N2 & Marino2013          & 8.42$\pm$0.20\\
O3N2 & Pérez-Montero2009   & 8.58$\pm$0.30\\
O3N2 & Pettini2004         & 8.56$\pm$0.31\\
\hline
mean O3N2 & & 8.60\\
std  O3N2 & & 0.12\\
\hline
mean N2+O3N2 & & 8.68\\
std  N2+O3N2 & & 0.14\\
\hline
\end{tabular}\\
\end{table}

The metallicity sensitive emission line ratios used in this work are the following:

\begin{equation*}\label{eq:2}
    N2 = \log(\frac{H\alpha}{[NII]\lambda6584})    
\end{equation*}

\begin{equation*}\label{eq:3}
    O3N2 = \log(\frac{[OIII]\lambda5007}{H\beta} \times \frac{H\alpha}{[NII]\lambda6584})   
\end{equation*}

Here H$\alpha$, H$\beta$, [NII]$\lambda$6584, [OIII]$\lambda$5007 represent the extinction corrected intrinsic fluxes of the respective spectral lines.  These metallicity indicators can be used for both, metal-poor and metal-rich galaxies in the range 7.31 $<$ 12 + log(O/H) $<$ 8.84 for the N2 and 
7.82 $<$ 12 + log(O/H) $<$ 8.78 for the O3N2 emission line ratios  on average. 
However, O3N2--based metallicity can reach values of 12 + log(O/H) $\leq$9.2 \citep{Naga06}.
\begin{table}
\centering
\caption{Observed F(H$\beta$) in units of 10$^{-17}$ erg/s/cm$^2$, de--reddened SDSS emission line fluxes, extinction c(H$\beta$) and emission line ratios.}
\label{tab:tab2}
\begin{tabular}{lccccccc}
\hline
                         & F($\lambda$)/F(H$\beta$) & I($\lambda$)/I(H$\beta$) \\
\hline
H$\beta$                 &  1.00$\pm$2.96 &  1.00$\pm$2.95 \\
$[OIII]\lambda$4959      &  $\dots$ & $\dots$ \\
$[OIII]\lambda$5007      &  1.29$\pm$2.69  &  1.18$\pm$2.48 \\
H$\alpha$                &  5.76$\pm$12.13 &  2.86$\pm$8.00 \\
$[NII]\lambda$6584       &  2.05$\pm$4.51  &  1.01$\pm$2.91\\
$[SII]\lambda$6717       &  2.93$\pm$6.32  &  1.39$\pm$4.05 \\
$[SII]\lambda$6731       &  0.97$\pm$2.16  &  0.46$\pm$1.36 \\
\hline
F(H$\beta$)              &  5.75$\pm$0.46 \\
c(H$\beta$)              &  1.02$\pm$2.69 \\
\hline
log([OIII]$\lambda$5007/H$\beta$)             &  0.07$\pm$0.91 \\
log([NII]$\lambda$6584/H$\alpha$)             & -0.45$\pm$0.32 \\
log([SII]$\lambda\lambda$6717,6731/H$\alpha$) & -0.19$\pm$0.22 \\
\hline
\end{tabular}\\
\end{table}

The population spectral synthesis (PSS) code FADO \citep[Fitting Analysis using Diﬀerential evolution Optimization,][]{gomes17} was applied to the J2306 SDSS spectrum corrected for Galactic extinction based on  \cite{1998ApJ...500..525S} color excess E(B-V) map and the \cite{Cardelli1989} extinction law (CCM). The  observed emission line fluxes of J2306 were obtained from FADO.  We note that due to poor SNR (see Figure \ref{fig-spec}) the FADO fit  to the [OIII]$\lambda$5007 line failed. However that line was then re--fitted using our own python  line fitting code and the flux obtained was utilised to calculate  12 + log(O/H) for J2306. The [OIII]$\lambda$4959 emission line is unresolved.
We then corrected the intrinsic extinction for observed flux F($\lambda$) to determine the intrinsic emission--line fluxes I($\lambda$)  (see Table \ref{tab:tab2}) relative to H$\beta$ using the formula:

\begin{equation*}\label{eq:4}
\frac{I(\lambda)}{I(H\beta)} = \frac{F(\lambda)}{F(H\beta)} \times
10^{c({H\beta})f(\lambda)}
\end{equation*}

where f($\lambda$) is the reddening function given by CCM assuming Rv = 3.1, c({H$\beta$}) is logarithmic reddening parameter calculated for case B, assuming Balmer decrement ratio H$\alpha$/H$\beta$ = 2.86 at 10 000 K, from \cite{2006agna.book.....O}. 

The 12 + log(O/H) mean of all the Table \ref{tab:tab1} calibrators was 8.68 with a standard deviation of 0.14 dex. Compared to other dwarf galaxies in the literature \cite[e.g.,][]{lee03}, our 12 + log(O/H) value is high and  approximately solar \citep[12 + log(O/H) = 8.69;][]{Asplund2009}, within the uncertainties, and is comparable to that of TDGs in the literature (see Figure \ref{fig1}).
We summarise the properties of J2306 in Table \ref{table3}

\begin{table} 
\centering
\caption{Properties of J2306}
\label{table3}

\begin{tabular}{lcc}
    \hline
    Property                   & Value                  &  Unit \\
     \hline
   
   Redshift                    &0.005                     & \\
   Distance                   &22.7                      &Mpc\\
   g band magnitude    &-14.83                   &mag\\
    g -- r                            &0.23\tablefootnote{The galaxy has an SDSS unreliable flag on its photometry. The error on the colour is $\sim$0.12. Therefore we  treat the stellar mass, estimated using this colour, as an order of magnitude estimate.}                      &mag \\
   12 + log(O/H)           & 8.68                      & \\
   $\rm D_{25}           $& 2.9                     &kpc\\
   Stellar mass             &2.4 $\times10^7      $&$\rm{M}_{\odot}$\\
   \hi\ mass                   &4.3$\times10^8    $&$\rm{M}_{\odot}$\\
   Dynamical mass     &6.4$\times10^9     $&$\rm{M}_{\odot}$\\
    \hline
\end{tabular}\\
\end{table}

\subsection{GMRT HI observations of J2306}
J2306  was observed in \hi\ at 21 cm using the GMRT on 30th May 2022,  with the pointing centre  at the projected position of J2306. A 12.5 MHz bandwidth was used giving a channel resolution $\sim$ 6 kHz. The observation details are listed in Table \ref{table4}. The \hi\ data \textcolor{black}{was} analysed using standard reduction procedures with the  Astronomical Image Processing System ({\tt AIPS}) software package. The flux density scale used was from \cite{baars77} with uncertainties of $\sim$5 percent. After bad data due to RFI and faulty antennas were flagged, the data was calibrated and
continuum subtracted in the uv--domain.  The {\tt AIPS} task {\sc{imagr}} was then
used to convert the uv--domain data to three dimensional \hi\ image cubes. To study the \hi\  distribution in detail, image cubes with different spatial resolutions
were made by varying the uv limits and applying different `tapers' to the data.
Finally, the {\tt AIPS} task {\sc{momnt}} was used to create the integrated \hi\ images and the velocity field maps from the \hi\ image cubes.  Properties of the low and medium--resolution maps presented in this paper are given in Table \ref{table4}.

\begin{table}
\centering
\begin{minipage}{110mm}
\caption{GMRT observational and \hi\  map parameters}
\label{table4}
\begin{tabular}{ll}
\hline
Rest frequency & 1420.4057 MHz \\
Observation Date & 30th May, 2022\\
Integration time  & \textcolor{black}{9.0 hrs } \\
Primary beam & 24\arcmin ~at 1420.4057 MHz \\
Low--resolution (beam--FWHP)  & 38.9$^{\prime\prime}$ $\times$ 36.9$^{\prime\prime}$, PA = -18.0$^{\circ}$ \\
Medium--resolution beam--FWHP& 24.5$^{\prime\prime}$ $\times$ 21.5$^{\prime\prime}$, PA =  9.5$^{\circ}$ \\
rms per channel for low--resolution map & 3.0 mJy beam$^{-1}$  \\
rms per channel for medium--resolution map & 2.9 mJy beam$^{-1}$  \\
RA (pointing centre)&\textcolor{black}{ 23$^{\rm h}$ 06$^{\rm m}$ 15.0$^{\rm s}$  }\\
DEC (pointing centre)& \textcolor{black}{ 14$^\circ$ 39$^\prime$ 27.6$^{\prime\prime}$}\\

\hline
\end{tabular}
\end{minipage}
\end{table}


\section{Results}
\begin{figure}
\begin{center}
\includegraphics[width=12.8cm]{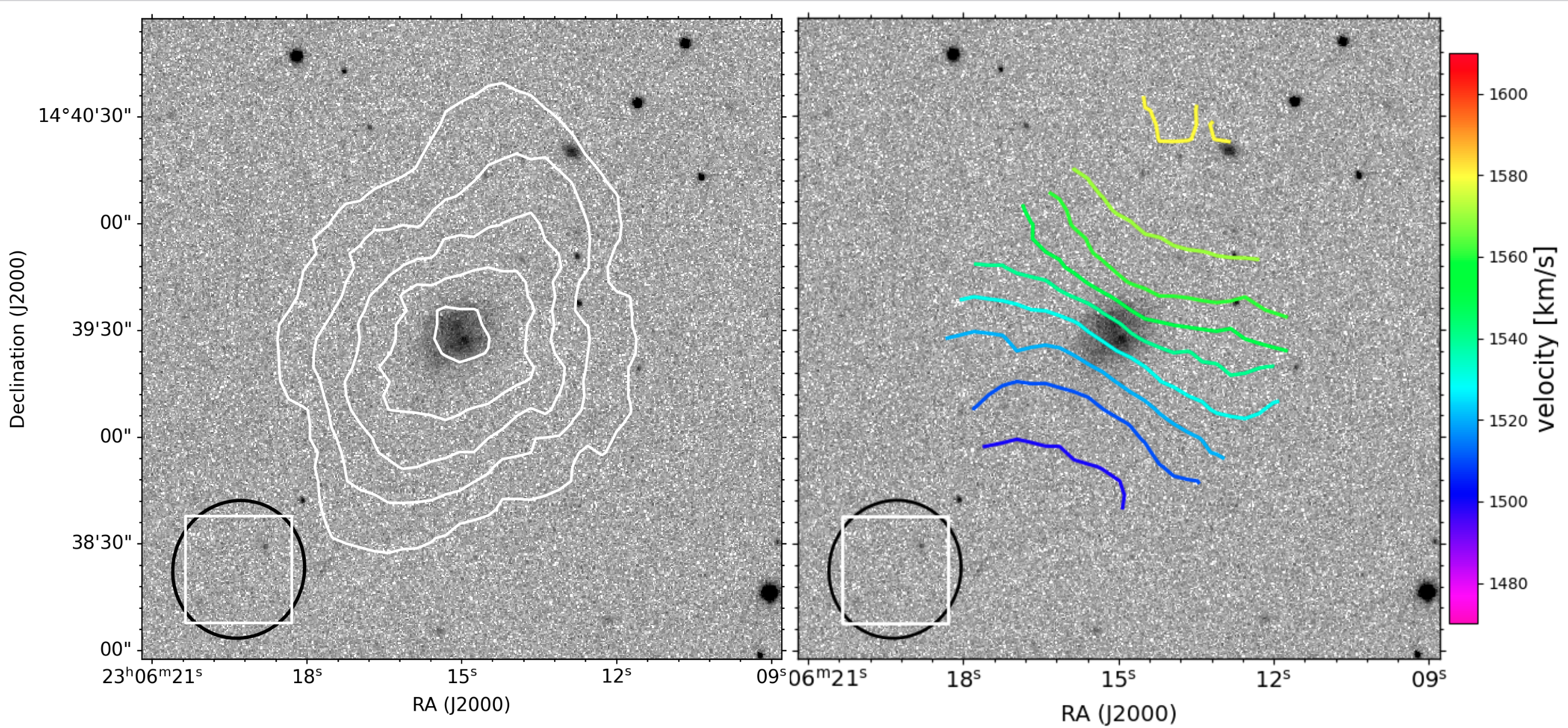} 
\caption{ \textcolor{black}{ J2306 low--resolution (38.9$^{\prime\prime}$ $\times$ 36.9$^{\prime\prime}$) GMRT moment maps. \textbf{Left:}  velocity integrated total \hi\ map  overlaid on SDSS g--band  image. The \hi\ column density contour levels are 10$^{19}$ \textcolor{black}{atoms} cm$^{-2}$ $\times$ (6.7, 15.7, 27.7, 33.7, 47.1). \textbf{Right:} velocity field of the galaxy overlaid on a SDSS g -- band  image.} The contours are from 1500 to 1580 \km, inclusive,  in steps of 10 \km\textcolor{black}{, per the colour scale}. The synthesised beam is plotted using a black ellipse at the bottom left of both the images. The white box on the ellipse denotes 30$^{\prime}$$\times$30$^{\prime}$ area.}
\label{fig4}
 \end{center}
\end{figure}

\begin{figure}
\begin{center}
\includegraphics[width=7.8cm]{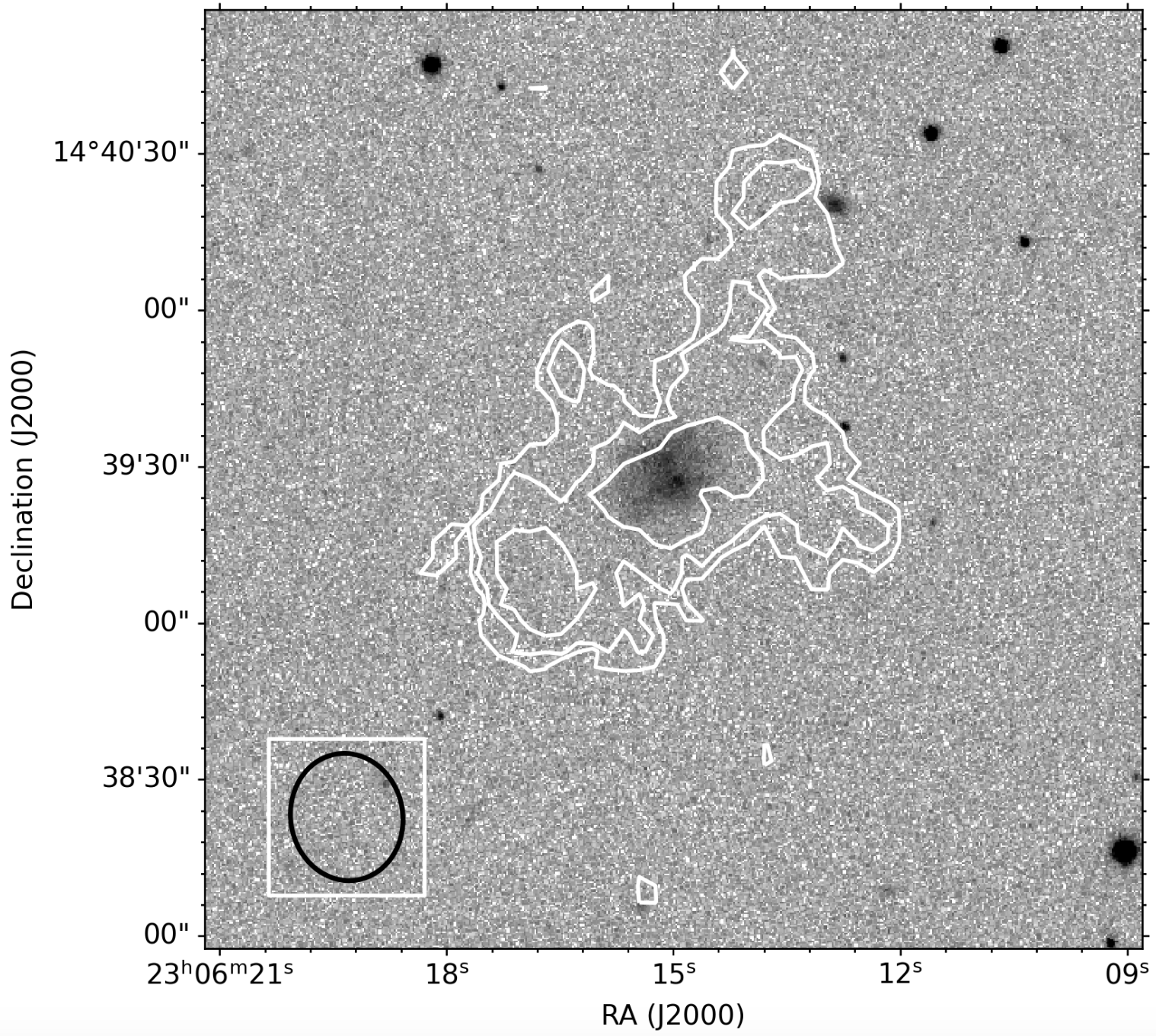} 
\caption{\textcolor{black}{J2306  medium--resolution (24.5$^{\prime\prime}$ $\times$ 21.5$^{\prime\prime}$) velocity integrated \hi\ map overlaid on SDSS g -- band} image of the galaxy. The \hi\ column density contour levels are 10$^{19}$ \textcolor{black}{atoms} cm$^{-2}$ $\times$ (5.2, 8.6, 17.3). The synthesised  beam is plotted using a black ellipse at the bottom left of the image. The white box on the ellipse denotes 30$^{\prime}$ $\times$ 30$^{\prime}$ area.}
\label{fig5}
 \end{center}
\end{figure}

The low--resolution GMRT integrated \hi\ and velocity field images reveal J2306's \hi\ disk to be extended and to a first order unperturbed.  The left panel of Figure \ref{fig4} shows the low--resolution (38.9$^{\prime\prime}$ $\times$ 36.9$^{\prime\prime}$) velocity integrated \hi\ flux density contours and the right panel shows the \hi\ velocity field for 
J2306, respectively. Figure \ref{fig5} shows the medium--resolution (24.5$^{\prime\prime}$ $\times$ 21.5$^{\prime\prime}$)  velocity integrated \hi\ flux density contours for  the galaxy.  At the distance of the galaxy, 22.7 Mpc, the 38.9$^{\prime\prime}$ and 24.5$^{\prime\prime}$ GMRT beams sample 4.3 kpc and 2.7 kpc, respectively. While the low--resolution \hi\ disk morphology appears highly symmetric and undisturbed, the medium--resolution image shows the highest column density \hi\ has a more irregular structure, with an overall alignment along a NW to SE axis. This is not unexpected, given that J2306 is a dwarf galaxy, and dwarfs are well known to have irregular \hi\ density distributions. The medium--resolution \hi\ column density maximum  spatially coincides with the optical galaxy.

The \textcolor{black}{low--resolution \hi\ velocity field shows a regularly rotating disk}. The velocity field maps are intensity weighted, and we present the low--resolution map here since it \textcolor{black}{has a higher} signal--to--noise ratio. \textcolor{black}{The regular rotation of the \hi\ disk is consistent with  the single dish Arecibo spectrum's}  double horn profile \textcolor{black}{which suggests a rotating \hi\ disk}. Resolved \hi\ imaging of the galaxy reveals the absence of any neighbour or companion galaxy within the GMRT 24$^{\prime}$ primary beam, \textcolor{black}{confirming} the entire  \hi\ mass detected with Arecibo belongs to J2306.

A comparison between the Arecibo and GMRT spectra of the galaxy confirms most of the \hi\ flux was recovered in the GMRT interferometric imaging. Usually for extended objects, flux loss occurs in the interferometric data due to a combination of several factors, namely flagging of crucial short baselines due to RFI, flagging of bad data leading to insufficient uv coverage and resolving out of the extended emission. Hence when available, it is preferable to carry out the global \hi\ estimates\textcolor{black}{, such as} \hi\ mass, using the single dish spectrum \textcolor{black}{data}. Of course  to use the single dish \hi\ flux to estimate a galaxy's \hi\ mass it is necessary to ensure that there is no contamination from other sources within  the single dish full width half power (FWHP) beam. In J2306's case the GMRT data allows us to confirm that there are no contaminating \hi\ sources within the single dish beam.
Using the integrated flux density from the Arecibo spectrum (3.54 Jy \km), we estimate the \hi\ mass of J2306 using Equation \ref{eq:2}:

\begin{equation}\label{eq:2}
     M(HI)=2.36 \times 10^{5} \times D^{2}\times \int S_{\nu} dv
\end{equation}

Here, D is the distance to J2306 \textcolor{black}{(22.7 Mpc)} and $\int S_{\nu} dv$ is the integrated flux density from the Arecibo spectrum. The \hi\ mass of J2306, thus calculated is $\sim$ 4.3$\times$10$^{8}$ \msolar. This implies the galaxy's M$_{HI}$/M$_{*}$   $\sim$18.   Even taking into account the caveat about the galaxy's stellar mass in Section \ref{source_props}   these ratios indicate that J2306  is an \hi\ rich dwarf galaxy.  Here M$_{*}$ and D$_{25}$  were estimated from the SDSS photometric data and NED respectively  and their values were used in compiling Table \ref{table3}. 



\section{Discussion}
Since the dwarf galaxy population is dominated by low metallicity objects \citep[e.g.,][]{lee03,kunth00}, the rarely found high--metallicity dwarfs are of particular interest \citep{peeples08}. One pathway to a high--metallicity dwarf is a result of tidal interactions between larger gas rich galaxies. During tidal interactions between such galaxies, gas rich tidal debris can form TDGs (M$_*$ $\sim$  10$^7$ -- 10$^8$ M$_{\odot}$). The collapsing metal rich gas debris is predicted to lead to rotating \hi\ and molecular disks and in turn in--situ star formation (SF) in the TDGs. Being born of tidal debris, TDGs are expected to have significantly higher metallicities and little or no DM content compared to normal dwarfs \citep{duc99,duc2014,2010AJ....140.2124B}. Another possible pathway to high--metallicity dwarf could be the accretion of nearby high--metallicity companions. Given the small size of J2306, any recently  accreted companion would have been an even smaller metal rich dwarf galaxy. Here we discuss the environment of J2306, estimate its dynamical mass and explore  possible scenarios under which  it could have attained such high metal abundance.

To understand the environment of J2306, we searched for neighbouring galaxies using the NED database. At  J2306's  distance, 60$^{\prime}$ is $\sim$ 380 kpc. Thus a 60$^{\prime}$ radius search, with an optical velocity (V$_{opt}$) constraint of 200 \km\ $\le$ V$_{opt}$ $\le$ 3500 \km\ was carried out using the NED database to search for neighbouring galaxies. This search returned only three neighbours. One of them is HIPASS J2306+14, which is the HIPASS \hi\  detection for J2306. 
There is a $\sim$ 4$^{\prime}$ difference between the HIPASS and WISE detected positions.
However, the velocities and the HIPASS spectrum reveal them to be the same galaxy.
The other two neighbours are AGES J230511+140404  and SDSS J230511.15+140345.7 at a projected distance of $\sim$ 250 kpc from J2306. 
These two detections are at same redshifts and separated by less than 0.5 $^{\prime}$  and \textcolor{black}{are almost certainly}  detections of the same galaxy, with the magnitude of \textcolor{black}{positional offsets}  within  the large positional uncertainties of the \textcolor{black}{Arecibo Galaxy Environment Survey (AGES)} survey FWHP beam (3.5 $^{\prime}$ Arecibo beam). 
 It is difficult to identify and confirm isolated galaxies. The standard isolation criteria require no companions within a projected sky diameter of $\sim$ 1Mpc \citep{verley07b}. In that respect, J2306  can be considered a fairly isolated galaxy with only one galaxy within a projected diameter $\sim$ 500 kpc and a velocity range of 200 -- 3500 \km. The relative isolation and the absence of obvious parent galaxy pair to generate  the tidal debris from which to form a TDG makes the recent TDG formation scenario for J2306 highly unlikely. 

\textcolor{black}{This raises the} interesting question of how, as an almost isolated dwarf galaxy, J2306 came to have  a higher than average metallicity. 
 SDSS J230511.15+140345.7 is the only galaxy with known redshift, within a diameter of 500 kpc of J2306. We studied its properties to see if this galaxy could have any effect on J2306's evolution. The galaxy is also a small dwarf galaxy, with an r -- band D$_{25}$ $\sim$ 0.3 kpc (extracted from NED),  g -- r colour $\sim$ 0.25.  This galaxy was detected in the AGES  and  it's \hi\ mass according to AGES Survey data is 1.2$\times$ 10$^{8}$ \msolar\ \citep{ages}. The optical size and the \hi\ mass indicates that SDSS J230511.15+140345.7  is also a dwarf galaxy. SDSS J230511.15+140345.7  is separated from J2306  by a projected distance of 250 kpc. We estimated the timescale of a hypothetical past interaction  between SDSS J230511.15+140345.7 and J2306. Considering their current positions, if both the galaxies moved apart at $\sim$ 200  \km\ (an average group dispersion velocity), the minimum distance covered by each of them would be 125 kpc. Since 250 kpc is their current projected separation, this suggests any past interaction would have taken place at least  6.3$\times$ 10$^{8}$ years ago. Such an interaction could in principle have increased the star formation and thereby enriched the gas in J2306. However, this timescale is too long for detectable signatures of interaction to remain identifiable in the \hi\ disk of J2306, and thus we cannot draw any conclusion about this possibility. Even for massive spiral galaxies, signs of past perturbations or mergers are identifiable in their disks only for about 4 -- 7 $\times$ 10$^{8}$ years \citep{holwerda11}. Interestingly, this timescale is sufficient to enrich the ISM of J2306.  Under such an interaction scenario the newly produced metals \textcolor{black}{would be} dispersed and mixed \textcolor{black}{on} $\sim$1 - 2 kpc scales in around 10$^{8}$ yrs \citep[e.g.,][]{tenorio96,lagos2009,lagos2016}.  One however cannot rule out the possibility of minor interactions  or mergers of even lower mass metal-rich, early--type dwarfs  surrounding J2306. More detailed spectroscopic data for J2306 might provide an answer about the feasibility of this option.

As discussed before, there are two well accepted criteria to identify a tidal dwarf galaxy, i.e., their higher than average metal abundance and their lack of DM.  J2306  is a high--metallicity dwarf with an \hi\ disk showing signs of regular rotation. We did not find any obvious progenitor galaxy pair near the dwarf galaxy. However, that cannot rule out that it is an old detached TDG which may have drifted away from its parent galaxy pair. We, therefore, estimated its dynamical mass to see if the galaxy is DM dominated or show signs of DM deficiency, within the measured \hi\ radius.   The ALFALFA Arecibo database shows the \hi\ line width (W$_{20}$) to be $\sim$106 \km. From the low--resolution GMRT \hi\ map we estimated the approximate \hi\ diameter of J2306  to be $\sim$13.2 kpc and the inclination to be $\sim$54$^{\circ}$. Using the inclination and the W$_{20}$ value we estimated the inclination corrected rotation velocity (V$_{rot}$) $\sim$ 63 \km\ and the dynamical mass (M$_{dyn}$) of the galaxy to be $\sim$6.4 $\times$ 10$^{9}$ \msolar. Comparing the M$_{dyn}$ to the baryonic mass 
\[\frac{\text{M$_{dyn}$}}{\text{M$_{gas}$$ + $M$_*$}} =9.4\], 

where M$_*$ = 2.4 $\times$ 10$^7$ \msolar, M$_{gas}$ = 65.2 $\times$ 10$^7$ \msolar, i.e. M$_{gas}$ (molecular + atomic) = 1.4 $\times$ M$_{HI}$ (43.0 $\times$ 10$^7$ \msolar).  Even allowing for the uncertainty in our estimated stellar mass, this ratio implies, like most regular dwarf galaxies, the galaxy is strongly DM dominated within its \hi\ radius. For TDGs this ratio would typically be closer to 1 \citep{seng14, sengupta2017}. This reinforces our previous conclusion, based on its environment, that J2306  is \textcolor{black}{almost certainly not} a tidal dwarf galaxy. 

To summarise,  we explored some possible reasons that could account for J2306's high gas metallicity. High--metallicity is a signature of TDGs so we checked for signs that J2306 is a TDG. The lack of a nearby progenitor pair and the dominance of DM within the  \hi\ radius rule out the galaxy as a TDG. Additionally, we did not find any signs of recent accretion or merger that could be an alternative explanation for the high gas \textcolor{black}{phase} metallicity.
 
\section{Conclusion}
We studied the high--metallicity dwarf galaxy J2306  to investigate whether  the galaxy could be a possible TDG candidate. The galaxy is not an early type dwarf, its g -- r colour is 0.23 mag and its optical morphology \textcolor{black}{is} non--compact and irregular. GMRT \hi\ mapping  of the galaxy confirmed it was  \hi\ rich and its unperturbed and rotating disk  extends $\sim$ 4 times further than the optical disk. We found no signs of past or ongoing interactions in the \hi\ images. Neither did we find any possible neighbouring galaxy pair which could potentially be the parent system if J2306 were to be a TDG.
 Using information from the \hi\ images, we estimated the dynamical mass of the galaxy. Contrary to what is expected for TDGs, J2306  was found to be a DM dominated galaxy. We explored other possibilities  (e.g., interactions or accretions) for the origin of the high--metallicity of J2306. However, we found J2306  to be fairly isolated with only one neighbouring galaxy within a projected diameter of 500 kpc. We conclude that while J2306 
 is a high--metallicity galaxy \textcolor{black}{this property is} neither of \textcolor{black}{recent} tidal origin, nor does it show any obvious signs of recent accretion or merger. It is located in a fairly isolated environment and thus its enrichment process could be a secular process\textcolor{black}{, or the result of an interaction in the distant past}. Further detailed spectroscopic observations of J2306 could provide an answer to how normal irregular dwarf galaxies  can achieve \textcolor{black}{such a level of metal} enrichment.

\section{Acknowledgements}

We thank the staff of the GMRT who have made these observations possible. The GMRT is operated by the National Centre for Radio Astrophysics of the Tata Institute of Fundamental Research.  YG acknowledges support from the National Key Research and Development Program of China (2022SKA0130100), and the National Natural Science Foundation of China (grant Nos. 12041306). PL (contract DL57/2016/CP1364/CT0010) and TS (contract DL57/2016/CP1364/CT0009) are supported by national funds
through Funda\c{c}\~{a}o para a Ci\^{e}ncia e a Tecnologia (FCT) and the Centro de Astrof\'isica da Universidade do Porto (CAUP). This research made use of APLpy, an open--source plotting package for Python \citep{Robitaille2012}. This research has made use of the Sloan Digital Sky Survey (SDSS). The SDSS Web Site is http://www.sdss.org/. This research has made use of the NASA/IPAC Extragalactic Database (NED) which is operated by the Jet Propulsion Laboratory, California Institute of Technology, under contract with the National Aeronautics and Space Administration.

\bibliographystyle{raa}
\bibliography{S189}
\end{document}